\documentclass[prl,reprint]{revtex4-1}%
\usepackage{amssymb}
\usepackage{amsfonts}
\usepackage{amsmath}
\usepackage{graphicx}%
\providecommand{\U}[1]{\protect\rule{.1in}{.1in}}

\begin{document}
\preprint{cond-mat/..}
\title[Short title for running header]{Coalescence model for crumpled globules formed in polymer collapse}
\author{Guy Bunin}
\affiliation{Massachusetts Institute of Technology, Department of Physics, Cambridge,
Massachusetts 02139, USA}
\author{Mehran Kardar}
\affiliation{Massachusetts Institute of Technology, Department of Physics, Cambridge,
Massachusetts 02139, USA}
\pacs{PACS number}

\begin{abstract}
The rapid collapse of a polymer, due to external forces or changes in solvent,
yields a long-lived `crumpled globule.' The conjectured fractal structure
shaped by hierarchical collapse dynamics has proved difficult to establish,
even with large simulations. To unravel this puzzle, we study a coarse-grained
model of in-falling spherical blobs that coalesce upon contact. Distances
between pairs of monomers are assigned upon their initial coalescence, and do
not `equilibrate' subsequently. Surprisingly, the model reproduces
quantitatively the dependence of distance on segment length, suggesting that
the slow approach to scaling is related to the wide distribution of blob sizes.

\end{abstract}
\volumeyear{year}
\volumenumber{number}
\issuenumber{number}
\eid{identifier}
\date[Date text]{date}
\received[Received text]{date}

\revised[Revised text]{date}

\accepted[Accepted text]{date}

\published[Published text]{date}

\startpage{1}
\endpage{ }
\maketitle

\begin{figure}
[ptb]
\begin{center}
\includegraphics[
height=2.6257in,
width=3.382in
]%
{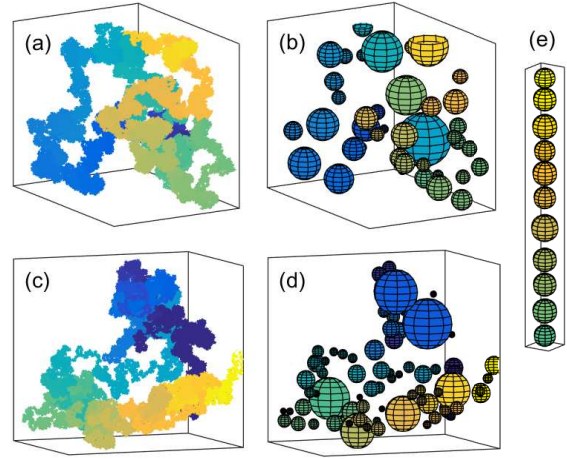}%
\caption{The Coalescence Model combines a simulation of coalescing spherical
drops (b,d,e), with estimates for distances between monomer pairs in final
structure. MD simulations (a,c) with initial conditions identical to (b,d) are
shown. Initial conditions are SAW for (a,b), RW for (c,d), and 1D for (e). All
panels show parts of the full systems. Colors show position of monomers along
the polymer, or average position for drops.}%
\label{fig:snapshots}%
\end{center}
\end{figure}

The rapid collapse of a polymer into a dense globule, is a long-standing
problem~\cite{Abrahams_EPL_collapse_dyn,stability_fract_glob_2,anomlous_diff_fract_glob,crooks_fractal_smoothing,deGennes,Grosberg_1988,Liberman_2009,Mirny_frac_glob_review,Pitard_Orland,raindrop,space_filling_curves,stability_fract_glob}%
. Such a collapse may be triggered by changes in solvent quality, causing the
polymer to reduce its solvent--exposed surface area by forming a dense
globule. A polymer may also be condensed by active forces, as in the
rearrangement of DNA by proteins in the cell nucleus~\cite{Liberman_2009}. The
rapid collapse does not allow sufficient time for formation of topological
entanglements which abound in an equilibrated compact
globule~\cite{Virnau,Grosberg_1988,Liberman_2009,Mirny_frac_glob_review,anomlous_diff_fract_glob,stability_fract_glob}%
. It is suggested~\cite{Grosberg_1988} that during collapse segments of the
polymer initially condense to sphere-like `blobs,' which coarsen upon contact
to form larger blobs. At any given time during the process the state is then
assumed to be characterized by a single length
scale~\cite{Grosberg_1988,Pitard_Orland,Abrahams_EPL_collapse_dyn,crooks_fractal_smoothing}%
; e.g., the typical size of the blobs or the width of the tube connecting the
blobs (see, e.g. Fig.~\ref{fig:snapshots}(a)~\cite{footnote_pearls}.) A
central assumption is that when two blobs coalesce they remain more or less
segregated within the newly formed structure. This is due to the slow
relaxation processes within blobs, and due to topological constraints which
prevent polymer segments forming the blobs from freely mixing -- unlike a melt
of independent polymer segments with open
ends~\cite{Grosberg_1988,topological_flory,topological_flory2,ring_meltPRL,rings_dont_mix1,rings_dont_mix2,Grosberg_ring_melt}%
. The final configuration is thus predicted to be a constant density,
self-similar, hierarchical structure, known as the `crumpled' or `fractal
globule'~\cite{anomlous_diff_fract_glob,Grosberg_1988,Liberman_2009,Mirny_frac_glob_review,raindrop,stability_fract_glob,stability_fract_glob_2}%
. The end-to-end distance $r_{m}$ of segments of length $m$ in the resulting
globule is predicted to scale as $r_{m}\sim m^{1/d}$ in $d$ space dimensions
(throughout the paper $d=3$). This is in contrast to the equilibrium state
reached at much later times \cite{deGennes}, where $r_{m}\sim m^{1/2}$ for
small $m$, saturating at the globule size $r_{\max}=N^{1/d}$, where $N$ is the
length of the polymer~\cite{footnote_p_contact}.

These predictions have been tested in several simulations of polymer
compaction~\cite{Liberman_2009,anomlous_diff_fract_glob,stability_fract_glob_2,stability_fract_glob,raindrop}%
, which generally confirm that the rapidly collapsed state is not entangled,
and indeed different from the equilibrium globule. However, they do not agree
upon its fractal nature. In particular, the expected scaling $r_{m}\sim
m^{1/d}$ has not been conclusively confirmed, even with the largest size
simulations (recently extended to polymers of up to 250,000
monomers~\cite{anomlous_diff_fract_glob,single_ring}). This implies either a
very large crossover scale before fractal behavior to clearly manifested, or
else that the collapsed state is not strictly self-similar. The large
finite-size effects are at least partly explained by partial mixing of short
polymer segments, for which the topological constraints do not
apply~\cite{Grosberg_1988,topological_flory,topological_flory2,ring_meltPRL,rings_dont_mix1,rings_dont_mix2,Grosberg_ring_melt,single_ring}%
. More generally, various protocols for constructing crumpled globules have
been
suggested~\cite{Liberman_2009,space_filling_curves,anomlous_diff_fract_glob,stability_fract_glob,stability_fract_glob_2,single_ring}%
, and it is not clear if the different procedures yield the same state.
Settling these issues seems to require even larger simulations, or new
theoretical insights.%

To this end, we propose a coarse-grained model for the crumpled globule formed
by polymer collapse, focusing here on active compression. The evolution of
blobs is modeled by drop coalescence, while distances between monomers in the
final crumpled state are assigned without keeping track of their individual
positions. This builds upon the topological segregation of blobs, and the
slowness of subsequent internal rearrangements, and should apply at scales
beyond which these conditions hold. The results highlight a different,
dynamical source of slow convergence to a self-similar state, and draw
connections to the physics of coagulation and drop-coalescence. In particular,
the assumption of single length-scale during the collapse may need to be
amended, at least when the collapse is due to active compression.

In spirit of the blob theory~\cite{deGennes}, the basic entity in our model is
a `drop,' an abstraction of the blob. A drop is a uniform density sphere which
contains a subset of the monomers from the original polymer, without explicit
\emph{position} assignments within the drop. Initially, every monomer is a
drop. Drops move (as detailed below) and coalesce into larger drops
immediately upon collision, i.e. as soon as the spheres overlap in space.
Drops do not break into smaller drops. Coalescence conserves volume; drops of
volume $v_{1},v_{2}$ forming a drop of volume $v_{1}+v_{2}$. (With the monomer
volume set to unity, the drop volume equals the number of monomers it
contains.) The new drop is centered at the center of mass of the coalesced
drops. The process terminates when all drops have merged to a single sphere.
Example coalescence runs are compared with corresponding molecular dynamics
(MD) runs in Fig.~\ref{fig:snapshots}.

Our main interest is the structure of the final collapsed state, as captured
by the distances $\{r_{m}\}$. However, in the coarse-grained drop coalescence
model we do not keep track of the internal structure of a drop. Instead we
assign \emph{distances-estimates} to \emph{all pairs of monomers} within a
drop. Guided by the slow internal rearrangements in the blob picture, a
distance-estimate is assigned when a pair of monomers first comes together in
a coalescence event, and is not changed thereafter. Upon coalescence of drops
of volumes $v_{1}$ and $v_{2}$, $v_{1}\times v_{2}$ pairs of monomers, one
from each drop, are assigned a distance-estimate. This distance is of order
$\left(  v_{1}+v_{2}\right)  ^{1/d}$, the linear size of the new drop, which
can be sampled from any distribution. As the results are highly insensitive to
this choice, we simply assign all $v_{1}v_{2}$ pairs the `distance' $\left(
v_{1}+v_{2}\right)  ^{1/d}$. The distance-estimates satisfy triangle
inequalities, and importantly are ultrametric, the tree structure reflecting
the hierarchy of the collapse process~\cite{Nechaev_stat_model_frac_globule}.
At the end, when all monomers belong to a single drop, all monomer pairs have
been assigned a distance-estimate.

To fully define the model, it remains to prescribe the drops' motion between
coalescence events, as well as the initial conditions. Here we focus on a
simple dynamics for active compression in which the drop velocity is
proportional to its distance from the origin, i.e. $d\tilde{R}_{\alpha
}/dt=-\tilde{R}_{\alpha}$, where $\tilde{R}_{\alpha}$ is the position of the
drop center. This corresponds to over-damped motion in a harmonic potential.
It can also be viewed as a uniform compression of space, as the distance
$\tilde{R}_{\alpha\beta}=\tilde{R}_{\alpha}-\tilde{R}_{\beta}$ between two
drops that have not yet coalesced changes as $d\tilde{R}_{\alpha\beta
}/dt=-\tilde{R}_{\alpha\beta}$, so all relative distances shrink by the same
factor per unit time. Apart from coalescence, there is no additional
interaction between drops. In particular, we do not impose polymeric bond
interactions between sequential monomers.

The initial conditions can be any configuration of the monomer positions. Here
we use random-walk (RW), self-avoiding walk (SAW), and `one-dimensional' (1D)
initial conditions. For the latter, monomers are positioned along a line as
${R}_{j}=j\hat{x}+\eta_{j}$, where $\hat{x}$ is the unit vector in the
$x$-direction, and $\eta_{j}$ are independent Gaussian variables in $d$
dimensions with $\left\langle \eta_{j}\right\rangle =0,\left\langle \eta
_{j}^{2}\right\rangle =c^{2}$, $c$ a number of order of drop radius. Open
polymers are used in all initial conditions. The SAW initial-conditions are
arguably the most natural for a polymer in a good solvent, but other initial
conditions help in obtaining additional insight. In particular, 1D initial
conditions are used only in the coalescence process, as the strong
unidirectional compression will cause large internal rearrangements after
blobs have formed within MD.

The coalescence model is compared with MD simulations where the polymer is
composed of monomers with standard polymeric bond attraction forces $F_{nn}$,
and repulsion $F_{rep}$ between all monomer pairs, see Supplemental Material.
Together with the same external force as in the coalescence model, the
monomers evolve as $dR_{i}/dt=-R_{i}+F_{nn}+F_{rep}$\textbf{.} The MD runs are
terminated when the polymer size stops decreasing. In both the coalescence and
MD simulations we do not add noise to the dynamics, as the entire collapse
process is fast (of order $\ln N$, see below). Noise may further increase
rearrangement processes in the MD simulations, which we try to minimize here.
A long-lived, metastable state is obtained, due to the time-scale separation
between the fast collapse process and further rearrangements which are much
slower. (As noted above, other `crumpled globule' metastable states may also
exist
\cite{anomlous_diff_fract_glob,stability_fract_glob,stability_fract_glob_2,single_ring}%
, at later times.)\textbf{ }%
\begin{figure}
[ptb]
\begin{center}
\includegraphics[
height=1.8128in,
width=3.7013in
]%
{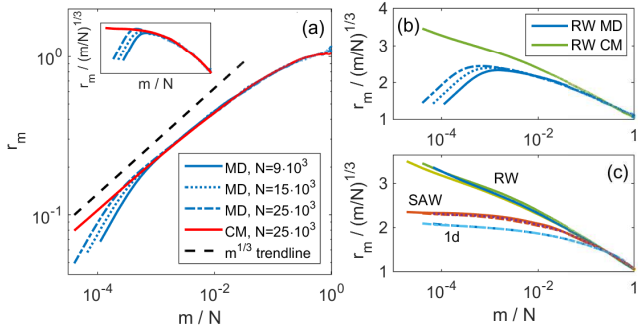}%
\caption{Comparison of Coalescence model and MD results. (a) Normalized
end-to-end distance $r_{m}$, with SAW initial conditions. Inset:
$r_{m}/\left(  m/N\right)  ^{1/3}$ for the same $r_{m}$. (b) $r_{m}/\left(
m/N\right)  ^{1/3}$ for RW initial conditions. (c) $r_{m}/\left(  m/N\right)
^{1/3}$ for Coalescence model with different initial conditions, and system
sizes, $N$.}%
\label{fig:r_m_compare_v2}%
\end{center}
\end{figure}

Now, let $d_{ij}^{coal}$ be the distance-estimate between monomers $i,j$ in
the coalescence model, and $d_{ij}^{MD}=\left\Vert R_{i}-R_{j}\right\Vert $
the corresponding distance between the $i,j$ monomers in the final state of
the MD run. In what follows we apply the same analysis to $d_{ij}%
^{coal},d_{ij}^{MD}$, refered jointly as $d_{ij}$. We define the normalized
end-to-end distance $r_{m}\equiv C\sqrt{\langle d_{ij}^{2}\rangle_{\left\vert
i-j\right\vert =m}}$, where the average $\left\langle \cdots\right\rangle
_{\left\vert i-j\right\vert =m}$ is over all monomer pairs separated along the
polymer by $m$ monomers, as well as over repeated runs of the two models, with
initial conditions resampled for each run. The normalization $C$ is chosen so
that $\sum_{m}\frac{N-m}{N\left(  N-1\right)  /2}r_{m}^{2}=1$. ($N-m$ is
simply the number of pairs that are $m$ apart in an open polymer of length
$N$). In this way the MD and coalescence models can be compared without any
fitting parameters.

Figure~\ref{fig:r_m_compare_v2}(a) compares $r_{m}$ from the coalescence model
with MD simulations for SAW initial conditions. MD results are shown for
different polymer lengths $N$. Strikingly, the MD and coalescence models agree
quantitatively over the entire range where the MD results of different polymer
sizes collapse, for $10^{-3}\lesssim m/N\leq1$. In the small $m$ regime, where
MD simulations with different $N$ do not collapse, they show a different
behavior (consistent with a random walk) not present in the coalescence model.
This demonstrates that the coalescence process indeed represents a
coarse-grained model, capturing large-scale behavior while incorporating
different microscopic details. The absence of polymeric bonds in the
coalescence model removes short-scale constraints that (as supported by the MD
simulations) do not effect large-scale properties through the distance assignment.

The $m^{1/3}$ trend-line in Fig.~\ref{fig:r_m_compare_v2}(a) indicates that
the expected scaling is not present at the tested system sizes $N\leq
25\times10^{3}$. To more carefully assess self-similarity, we study
finite-size scaling in the standard form
\begin{equation}
r_{m}=\left(  m/N\right)  ^{1/3}f\left(  m/N\right)  \ ,
\end{equation}
expected to hold for $1\ll m$. (The additional $N^{-1/3}$ reflects the choice
of normalization.) The scaling function $f\left(  x\right)  $ should go to a
constant for $x\rightarrow0$, such that $r_{m}\propto\left(  m/N\right)
^{1/3}$ for $1\ll m\ll N$. Data for both MD and coalescence models, for
different initial conditions and system sizes, are summarized in
Fig.~\ref{fig:r_m_compare_v2}. The difference between MD and coalescence
results for the RW initial conditions is larger than for the SAW. Importantly,
for both SAW and RW initial conditions, and for both models, the expected
condition, $f\left(  m/N\rightarrow0\right)  =const$, is not seen clearly for
tested values of $N$; in all cases the maximum of $f\left(  m/N\right)
=r_{m}/\left(  m/N\right)  ^{1/3}$ appears to grow with increasing $N$. This
effect is almost absent for the 1D initial conditions, and largest in the RW
case. Since the coalescence model includes only a minimal set of microscopic
details, it is surprising to see such a slow approach to the expected scaling
behavior. Understanding this trend in the coalescence model should provide
insights into the more complex case of the collapsing polymer.

While simplified, a full understanding of the coalescence model -- including
the distribution of distance-estimates assigned as a function of the time $t$
and separation $m$ -- is still difficult. Fortunately, some insights regarding
scaling (or lack thereof) can be gained by examining the distribution of drop
volumes as a function of time, even without making reference to the
distance-estimate assignments. In particular, the volume distributions already
show a slow approach to scaling.

Let $\rho_{t}\left(  v\right)  $ be the distribution of drop sizes at time
$t$. Volume conservation implies that $\int dvv\rho_{t}\left(  v\right)  =N$,
while $\int dv\rho_{t}\left(  v\right)  $ is the number of drops at time $t$
(averaged over repeated runs). Within the scaling picture, the dynamics of the
collapse is described by a single typical drop size as a function of time,
$\bar{v}\left(  t\right)  $, when $1\ll\bar{v}\left(  t\right)  \ll N$. For
example, in the tube picture $\bar{v}^{1/d}$ may be the thickness of the tube.
In such a scaling regime, $\rho_{t}\left(  v\right)  $ should depend on time
only through $\bar{v}\left(  t\right)  $, as
\begin{equation}
\rho_{t}\left(  v\right)  =\frac{N}{\bar{v}\left(  t\right)  ^{2}}\phi\left(
\frac{v}{\bar{v}\left(  t\right)  }\right)  \ . \label{eq:coal_vol_scaling}%
\end{equation}
The factor $\bar{v}^{-2}$ ensures that $\int dvv\rho_{t}\left(  v\right)  $
remains constant. We measure the typical size via $\bar{v}\left(  t\right)
\equiv\left\langle \sum_{n}v_{n}^{2}\right\rangle /\sum_{n}v_{n}=\left\langle
\sum_{n}v_{n}^{2}\right\rangle /N$, where the sums run over all drops at time
$t$, and the average \thinspace$\left\langle \cdots\right\rangle $ is over
initial conditions. This definition, standard in drop coalescence/coagulation
literature~\cite{Leyvraz_review,Dongen_Ernst,Family_Meakin_PRL,Family_Meakin_PRA}%
, addresses possible divergences of $\rho_{t}\left(  v\right)  $ at small
volumes (see below). To test Eq.~\ref{eq:coal_vol_scaling}, we plot $\phi
_{t}\left(  v\right)  \equiv N^{-1}\bar{v}\left(  t\right)  ^{2}\rho
_{t}\left(  v\right)  $, against the normalized volume\ $v/\bar{v}\left(
t\right)  $, and check for data collapse at different values of $t$.%
\begin{figure}
[ptb]
\begin{center}
\includegraphics[
trim=0.306468in 0.000000in 0.307477in 0.000000in,
height=2.3593in,
width=3.5863in
]%
{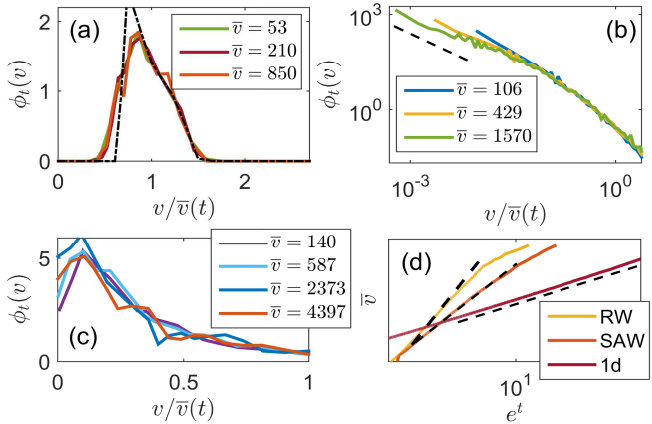}%
\caption{Drop volume distributions in the Coalescence model. (a) 1d initial
conditions. Dashed line: sequential collisions theory. (b) RW initial
conditions, dashed line: $x^{-1}$ trendline. (c) SAW initial conditions.
(d)\ $\bar{v}(t)$ for the three models. Dashed lines: $e^{wt}$ for different
cases (see text).}%
\label{fig:vol_dist_all}%
\end{center}
\end{figure}

The distributions, depicted in Figs.~\ref{fig:vol_dist_all}(a,b) for 1D and RW
initial conditions, respectively, are quite different. The one in (a) (1D
initial conditions) is concentrated in the region $0.4\lesssim v/\bar
{v}\lesssim1.6$, and strongly suppressed outside this interval. Thus, at any
given time all volumes are of the same order, with a ratio of about 4 between
the largest and smallest. The distributions at different times collapse nicely
when plotted against $v/\bar{v}\left(  t\right)  $. In contrast, the
distributions for RW initial conditions in Fig.~\ref{fig:vol_dist_all}(b) are
very wide (note the log-log scale), with possibly a diverging tail at small
volumes, $v/\bar{v}\ll1$. (The dashed line, $x^{-1}$, is included as an
indication of such potential divergence.) Moreover, the distributions fail to
collapse in this tail. With SAW initial conditions,
Figs.~\ref{fig:vol_dist_all}(c), the results appear to be intermediate between
the above two cases, with a distribution that is finite at $v=0$ (at least for
tested $N$). The simulations in Figs.~\ref{fig:vol_dist_all}(a,b,c) were
carried out with $N=2.5\times10^{4}$, $5\times10^{4}$, and $10^{5}$
respectively, to allow for better scaling in Figs.~\ref{fig:vol_dist_all}(b,c).

The evolution of $\bar{v}\left(  t\right)  $ is depicted in
Fig.~\ref{fig:vol_dist_all}(d). For 1D and SAW initial conditions $\bar
{v}\left(  t\right)  \propto e^{wt}$ is a good fit at intermediate times,
where $w=\frac{d}{d\nu_{0}-1}$, with $\nu_{0}$ describing the scaling
$r_{m}^{\left(  0\right)  }\sim m^{\nu_{0}}$ in the initial condition. This
form is explained as follows: At time $t$, segments of initial length $\bar
{v}\left(  t\right)  $ form drops of volume $\bar{v}\left(  t\right)  $, with
diameter $\left[  \bar{v}\left(  t\right)  \right]  ^{1/d}$. The initial
end-to-end distance of these segments scales as $\left[  \bar{v}\left(
t\right)  \right]  ^{\nu_{0}}$. If this reduction in length follows the
general exponential approach of free monomers~\cite{footnote_exponential},
then $\left[  \bar{v}\left(  t\right)  \right]  ^{\nu_{0}}e^{-t}=\left[
\bar{v}\left(  t\right)  \right]  ^{1/d}$, leading to the above
relation~\cite{footnote_dynamic_exp}. For RW initial conditions the growth of
$\bar{v}\left(  t\right)  $ follows this form for a narrower interval, with
wide crossover regions, probably related to the lack of scaling observed in
Fig.~\ref{fig:vol_dist_all}(b).

The difference between the three volume distributions in
Fig.~\ref{fig:vol_dist_all} can be directly observed in the coalescence and MD
runs in Fig.~\ref{fig:snapshots}; note especially the large number of small
drops for RW initial conditions in Fig.~\ref{fig:snapshots}(d). In the MD
starting from the same initial conditions, these manifest as open segments of
the polymer, alongside large condensed regions, see Fig.~\ref{fig:snapshots}%
(c); the thickness of a putative `tube' is highly uneven. We now present
theoretical approaches to analogous problems leading to the two very different
volume distributions obtained in Figs.~\ref{fig:vol_dist_all}(a,b).

Broad distributions, with power-law tails at small volumes, as in
Fig.~\ref{fig:vol_dist_all}(b) for RW initial conditions, appear in related
problems such as diffusion-limited aggregation~\cite{DLA_cluster_survival},
and drop coalescence~\cite{Family_Meakin_PRL,Family_Meakin_PRA}. Heterogeneous
drop coalescence, where drops grow but no new drops are added, is perhaps most
similar to our model. There, the distribution of drop sizes is highly
poly-disperse when $\bar{v}\left(  t\right)  $ (in our notation) grows
exponentially in time~\cite{Family_Meakin_PRA}. Aggregating systems are
commonly studied through the (mean-field) Smoluchowski
equation~\cite{Dongen_Ernst,Leyvraz_review}:
\begin{equation}
\partial_{t}\rho\left(  v\right)  =\frac{1}{2}\int dv^{\prime}J_{t;v^{\prime
},v-v^{\prime}}-\int dv^{\prime}J_{t;v^{\prime},v}\ .\label{eq:Smoloch}%
\end{equation}
The first term is the change in density $\rho_{t}\left(  v\right)  $ due to
creation of drops of size $v$, while the second term describes their removal
due to coalescence. In the current context, this approximation postulates a
rate $J_{t;v_{1},v_{2}}=K_{v_{1},v_{2}}\rho_{t}\left(  v_{1}\right)  \rho
_{t}\left(  v_{2}\right)  $ for collisions between drops of volumes
$v_{1},v_{2}$, with a kernel $K_{v_{1},v_{2}}$ depending only on the volumes
of the coalescing drops. Here, rather than explicitly constructing an
approximate kernel, we refer to extensively studied scaling solutions of the
Smoluchowski equation~\cite{Dongen_Ernst,Leyvraz_review}. In particular, for
homogeneous kernels such that $K_{av_{1},av_{2}}=a^{\lambda}K_{v_{1},v_{2}}$
for $a>0$, it is known that when $\bar{v}\left(  t\right)  \propto e^{wt}$, as
in our case, the scaling function $\phi\left(  x\right)  $ in
Eq.~\ref{eq:coal_vol_scaling} has the following properties: It is strongly
(exponentially) decaying for $x\gg1$, and has a diverging power-law tail
$\phi\left(  x\right)  \sim x^{-\tau}$ for $1\gg x$, with $1\leq\tau<2$ (the
value of $\tau$ depends on the kernel). For example, the kernel $K_{v_{1}%
,v_{2}}=v_{1}+v_{2}$ admits an exact scaling solution $\phi\left(  x\right)
=\frac{1}{\sqrt{2\pi}}x^{-3/2}e^{-x/2}$ with $\bar{v}=e^{2t}$, so that
$\tau=3/2$.

Interestingly, a slow approach to the asymptotic scaling is well documented
for certain cases with such small-volume
tails~\cite{Kang_crossover,Smol_init_cond,Leyvraz_review,Breath_fig_low_tails_collapse}%
. The nature of this slow approach is still not well-understood, and might be
sensitive to the kernel form and initial conditions. It is intriguing to
speculate on its relation to the present problem.

In the case of 1D initial conditions in Fig.~\ref{fig:vol_dist_all}(a),
essentially all drops at a given time have volumes of the order of $\bar
{v}\left(  t\right)  $. Unlike RW initial conditions, all collisions here are
sequential, i.e. between drops comprised of adjacent segments along the
polymer. Here geometry matters: after a collision gaps are formed on both
sides of the newly constructed drop, greatly reducing its probability of
coalescing again before other drops have time to grow. Smaller drops leave
smaller gaps, and have an increased probability of additional collisions.
These effects are discussed in a quantitative way in the Supplemental
Material. An approximate evolution equation is derived that admits the scaling
solution shown in Fig.~\ref{fig:vol_dist_all}(a) (dashed line), which strongly
decays outside a narrow interval of $v/\bar{v}$, just like the results from
the full coalescence model.

The coalescence model proposed here does away with several microscopic details
present in the collapsing polymer that could delay the asymptotic approach to
scaling. The lack of simple scaling in coalescence thus points to deeper
problems with the simple model of hierarchical collapse, in particular in the
assumption of a sharply defined blob-scale. The coalescence model is
interesting in its own right; it is closely related to widely studied problems
of coarsening of growing water drops, but differs in the initial (polymeric)
distribution of droplets. As demonstrated, different such initial conditions
lead to widely dissimilar probability distributions. Probing the role of
fractal initial conditions in coalescence problems should thus be worthy of
further exploration.

We thank Bernard Derrida, Maxim Imakaev, Leonid Mirny and Adam Nahum for
valuable discussions. This research was supported by the NSF through grant No.
DMR-12-06323. GB acknowledges the support of the Pappalardo Fellowship in Physics.

\section{Supplemental Material}

\section{Molecular dynamics (MD)\ Simulation details}

In the MD simulations, the polymer is composed of monomers with standard
polymeric bond forces $F_{nn}$, and repulsion $F_{rep}$ between all monomer
pairs. $F_{nn},~F_{rep}$ are given by the gradients $F_{nn}=-\nabla
U_{nn},~F_{rep}=-\nabla U_{rep}$, where~\cite{MD_potentials}%
\begin{align*}
U_{nn}\left(  r\right)   &  =\left\{
\begin{array}
[c]{ccc}%
\frac{1}{2}kl_{\max}^{2}\ln\left[  1-\left(  r/l_{\max}\right)  ^{2}\right]  &
& r<l_{\max}\\
\infty &  & r>l_{\max}%
\end{array}
\,,\right. \\
U_{rep}\left(  r\right)   &  =\left\{
\begin{array}
[c]{ccc}%
4\varepsilon\left[  \left(  \sigma/r\right)  ^{12}-\left(  \sigma/r\right)
^{6}\right]  +C\left(  r_{cut}\right)  &  & r<r_{cut}\\
0 &  & r>r_{cut}%
\end{array}
\,,\right.
\end{align*}
with $k=30,~l_{max}=1.5,~\varepsilon=\sigma=1$~\cite{MD_potentials}, and
$r_{cut}=2^{1/6}$ (which makes $U_{rep}$ purely repulsive). $C\left(
r_{cut}\right)  $ is chosen such that $U_{rep}\left(  r=r_{cut}\right)  =0$.

Together with the same external force as in the coalescence model, the
monomers evolve as $dR_{i}/dt=-R_{i}+F_{nn}+F_{rep}$. We use first-order
integration in time. The MD runs are terminated when the polymer size stops
decreasing. Of course, there may be further internal rearrangements due to
reptation, but on much longer time-scales. The final states are nearly
spherical, with at most a few percent ellipticity.

\section{Theory for drop evolution on a line}

Here we further discuss the evolution of drop sizes for 1D initial conditions,
and describe a theory that accounts for the distribution of volumes in Fig.~3(a).

A discussed in the Main Text, for 1D initial conditions essentially all drops
at a given time have volumes of the order of $\bar{v}\left(  t\right)  $, see
Fig.~3(a). All collisions here are sequential, i.e. between drops comprised of
adjacent segments along the polymer. Here geometry matters: after a collision
gaps are formed on both sides of the newly constructed drop, greatly reducing
its probability of coalescing again before other drops have time to grow.
Smaller drops leave smaller gaps, and have an increased probability of
additional collisions.

To quantify this effect, note that a drop moves according to $R\left(
t\right)  =R_{0}e^{-t}$, where $R_{0}$ is the center of mass of the monomers'
initial positions. Consider two sequential segments, containing $v_{1}$ and
$v_{2}$ monomers respectively. If the segments are chosen at random (while
still being adjacent), the distance $\delta x_{0}$ between their centers is
Gaussian distributed, with average $\frac{v_{1}+v_{2}}{2}$ and variance
$c\left(  v_{1}+v_{2}\right)  $ (the average distance between neighboring
monomers is taken to be one). If these segments form drops, they will meet
when $\left(  {v_{1}^{1/d}+v_{2}^{1/d}}\right)  e^{t}\simeq\left(  \frac
{v_{1}+v_{2}}{2}\pm c\sqrt{v_{1}+v_{2}}\right)  $. The resulting time frame is
narrow when $v_{1,2}\gg1$. Small drops are rapidly swept away as they coalesce
with typical sized drops at earlier times than larger drops.

The above argument can be turned into an evolution equation for $\rho
_{t}\left(  v\right)  $. For other approaches to a related problem, see
\cite{Derrida_1d_EPL,Derrida_1d_long}. We postulate a collision rate
\begin{equation}
J_{t;v_{1},v_{2}}=K_{v_{1},v_{2};t}\rho_{t}\left(  v_{1}\right)  \rho
_{t}\left(  v_{2}\right)  \ , \tag{S1}%
\end{equation}
as in Equation~3 in the Main Text, but with a kernel $K_{v_{1},v_{2};t}$ that
explicitly depends on time (therefore this is not a Smoluchowski equation).
The kernel represents the probability per unit time for collision of drops
$v_{1},v_{2}$, given that they have not yet coalesced. It corresponds to
$\left(  \partial S_{2}/\partial t\right)  /S_{2}$, where $S_{2}(t;v_{1}%
,v_{2})$ is the survival probability of the two drops till time $t$. We
approximate $S_{2}$ by the probability that two drops formed by a random pair
of consecutive segments do not overlap at time $t$, i.e., $\delta x_{0}%
e^{-t}>v_{1}^{1/d}+v_{2}^{1/d}$,%
\begin{equation}
S_{2}(t;v_{1},v_{2})=\frac{1}{\sqrt{2\pi}c}\int_{G\left(  v_{1},v_{2}%
,t\right)  }^{\infty}e^{-\frac{\eta^{2}}{2c^{2}}}d\eta\tag{S2}%
\end{equation}
where
\begin{equation}
G\left(  v_{1},v_{2},t\right)  \equiv\frac{e^{t}\left(  v_{1}^{1/d}%
+v_{2}^{1/d}\right)  -\frac{v_{1}+v_{2}}{2}}{c}\ , \tag{S3}%
\end{equation}
so that the kernel takes the form%
\begin{equation}
K_{\left(  v_{1},v_{2},t\right)  }=\frac{e^{-\frac{G^{2}}{2}}\partial_{t}%
G}{\int_{G}^{\infty}e^{-\frac{\eta^{2}}{2}}d\eta}=\frac{e^{-\frac{G^{2}}{2}%
}\partial_{t}G}{\operatorname*{erfc}G}\ . \tag{S4}%
\end{equation}
Equation~3 with this kernel indeed has a scaling solution with $\bar{v}\propto
e^{wt}$, as can be verified by direct substitution. Numerically solving Eq.~3
with this kernel, one obtains a scaling collapse with $\bar{v}$ quickly
approaching $e^{wt}$, and a scaling form $\phi\left(  \frac{v}{\bar{v}%
}\right)  $ shown in Fig.~3(a) (dashed line), which strongly decays outside a
narrow interval of $v/\bar{v}$, just like the results from the full
coalescence model.


\begin{thebibliography}{99}
\bibitem {footnote_dynamic_exp}In terms of $\xi\equiv e^{-t}$, the linear
compression of space, $w$ is a dynamic scaling exponent $\bar{v}\sim\xi^{w}$,
similar to the one in coalescence models
\cite{Leyvraz_review,Dongen_Ernst,Family_Meakin_PRL,Family_Meakin_PRA}.
\end{thebibliography}
\end{document}